\documentclass[prl,aps,showpacs,twocolumn]{revtex4}
\usepackage{epsfig}
\usepackage{bm}

\begin{document}

\title{Fluctuations of temperature gradients in turbulent thermal convection}

\author{\small K.R. Sreenivasan$^{1}$, A. Bershadskii$^{1,2}$ and J.J. Niemela$^1$}
\affiliation{\small {\it $^1$International Center for Theoretical
Physics, Strada Costiera 11, I-34100 Trieste, Italy}\\
{\it $^2$ICAR, P.O. Box 31155, Jerusalem 91000, Israel}}

\begin{abstract}
Broad theoretical arguments are proposed to show, formally, that
the {\it magnitude} $G$ of the temperature {\it gradients} in
turbulent thermal convection at high Rayleigh numbers obeys the
same advection-diffusion equation that governs the temperature
fluctuation $T$, except that the velocity field in the new
equation is substantially smoothed. This smoothed field leads to a
$-1$ scaling of the spectrum of $G$ in the same range of scales
for which the spectral exponent of $T$ lies between $-7/5$ and
$-5/3$. This result is confirmed by measurements in a confined
container with cryogenic helium gas as the working fluid for
Rayleigh number $Ra=1.5\times 10^{11}$. Also confirmed is the
logarithmic form of the autocorrelation function of $G$. The
anomalous scaling of dissipation-like quantities of $T$ and $G$
are identical in the inertial range, showing that the analogy
between the two fields is quite deep.
\end{abstract}

\pacs{47.27.Te; 47.27.Jv}

\maketitle

While statistical properties of temperature fluctuations in
turbulent Rayleigh-B\'enard convection have received considerable
experimental and theoretical consideration (see, for instance,
\cite{lib}-\cite{ching} and the references therein), corresponding
properties of temperature gradients are still unexplored from both
theoretical and experimental points of view. In the present paper,
we study statistical properties of temperature gradients
emphasizing their qualitative and quantitative similarity to those
of the temperature fluctuations themselves. Theoretical
considerations will be based on an equation to be derived for the
{\it magnitude} of temperature gradients, and the results will be
compared with measurements in turbulent convection in a confined
container of circular cross-section. We would like to emphasize
crucial difference between spectrum of the temperature gradient
and spectrum of {\it absolute} value (magnitude) of the
temperature gradient. The former one can be readily estimated
using the Taylor's hypothesis. Namely, the Fourier transform of
the time series of the temperature derivative is simply
proportional to omega multiplied by the transform of the
temperature time series. If this is the case, there should simply
be a factor omega squared between the two spectra. If, however, we
consider spectrum of absolute value of the temperature gradient,
then there is no straightforward relation between this spectrum
and spectrum of the original temperature time series. In the last
case one need use of physics, and the spectrum (autocorrelation
function) of the absolute value of the temperature gradient can
give additional information about the thermal convection process.

The measurement apparatus has unity aspect ratio. The sidewalls of
the apparatus are insulated, and the bottom and top walls are
maintained at constant temperatures; the bottom wall is held at a
slightly higher temperature $\Delta$ than the top wall. The
working fluid is cryogenic helium gas. We measure temperature
fluctuations at various Rayleigh numbers towards the upper end of
this range, in which the convective motion is turbulent, but use
the data obtained at the Rayleigh number $Ra=1.5\times 10^{11}$ in
the present paper. Time traces of fluctuations are obtained at a
distance of 4.4 cm from the sidewall on the center plane of the
apparatus. This position is outside of the boundary layer region
for the Rayleigh number considered here. At this Rayleigh number,
the mean wind (which is the large-scale circulation within the
convection apparatus) is well developed so Taylor's hypothesis can
be employed when necessary. More details of the experimental
conditions and measurement procedure can be found in Ref.\
\cite{n}.

In thermal convection, the temperature field $T({\bf r},t)$ is
convected by the velocity field ${\bf v} ({\bf r},t)$, which
itself is generated by density differences set up between the top
and bottom walls. We will consider incompressible flow obeying
$\nabla\cdot{\bf v} = 0$ (with unit density for simplicity). The
relevant equations under the Boussinesq approximation are
$$
\frac{\partial {\bf v}}{\partial t}= - \nabla p -({\bf v}\cdot
\nabla) {\bf v} +\nu \nabla^2 {\bf v} +\alpha g T \hat{z},
\eqno{(1)}
$$
$$
\frac{\partial T}{\partial t}=-({\bf v}\cdot \nabla) T + D
\nabla^2 T.  \eqno{(2)}
$$
Here $p$, $\nu$, $D$, $\alpha$, $g$ and $\hat{z}$ are,
respectively, the pressure, kinematic viscosity, thermal
diffusivity, isobaric thermal expansion coefficient, acceleration
due to gravity, and the unit vector in the upward direction.
Equation (2) is the standard scalar advection-diffusion equation,
except that the velocity field is coupled to the temperature
field. This `active' nature of the temperature fluctuations in
convection makes their statistical properties different from those
of a {\it passive} scalar advected by a turbulent velocity with no
back reaction. Restricting attention, for simplicity, to Prandtl
numbers of the order unity, the experimentally measured spectral
density of temperature fluctuations in the inertial (convective)
range rolls off at a rate that is closer to $-1.4$ than to $-5/3$,
the latter being the case for passive scalars in three-dimensional
homogeneous turbulence \cite{lib,y,as,tong,n}.

The equation for temperature gradients ${\bf G} \equiv \nabla T$
can be readily derived from (2) as
$$
\frac{\partial G_i}{\partial t}=-v_j\frac{\partial G_i}{\partial
x_j} - \frac{\partial v_j}{\partial x_i}G_j + D \frac{\partial^2
G_i}{\partial x_j^2},  \eqno{(3)}
$$
with the indices $i$ and $j$ representing the space coordinates,
and the summation over repeated indexes is assumed. The magnitude
$G$ of the temperature gradient is determined by ${\bf G} = G {\bf
n}$, where ${\bf n}$ is the unit vector with its direction along
vector ${\bf G}$. Multiplying both sides of Eq.\ (3) by $n_i$,
making summation over $i$, and taking into account of the fact
that $n_i^2 =1$, we obtain
$$
\frac{\partial G}{\partial t}= -({\bf v}\cdot \nabla) G + D
\nabla^2 G - \lambda G, \eqno{(4)}
$$
which is formally similar to Eq.\ (2) except for the last term in
(4). The coefficient $\lambda$ in this term has the form
$$
\lambda = n_in_j \frac{\partial v_i}{\partial x_j} + D \left(
\frac{\partial n_i}{\partial x_j}\right)^2.  \eqno{(5)}
$$

Let us now search for circumstances under which the last term in
Eq.\ (4) is small in the inertial range. The second term in
$\lambda$ is assured to be small because the diffusivity $D$ is
small. But the nature of the ``stretching" part on the right hand
side of Eq.\ (5) is not apparent without further considerations.

As a further step, let us make the following conditional average
of Eq.\ (4). Fix the magnitude $G$ in the vector field ${\bf G} =
G{\bf n}$ while performing the average over all realizations of
the direction vector field ${\bf n}$ permitted by equation (3).
Let us denote this ensemble average as $\langle ... \rangle_{{\bf
n}}$. From the definition, this averaging procedure does not
affect $G$ itself, but modifies the velocity field ${\bf v}$,
which in turn modifies the coefficient $\lambda$ in Eq.\ (4). We
may write
$$
\frac{\partial G}{\partial t}= -(\langle {\bf v} \rangle_{{\bf n}}
\cdot \nabla) G + D \nabla^2 G + \langle \lambda \rangle_{{\bf
n}}G. \eqno{(6)}
$$
It is worth noting that the solutions of Eq.\ (3) satisfy Eqs.\
(4) and (6), but not all possible formal solutions of the Eqs.\
(4) and (6) satisfy Eq.\ (3); similarly, not all formal solutions
of Eq.\ (6) satisfy Eq.\ (4) while all solutions of Eq.\ (4) do
satisfy Eq.\ (6). In particular, the solutions of Eqs.\ (4) and
(6) are the same only if: (a) the initial conditions for the two
equations are the same, and (b) if realizations of $\langle {\bf
v} \rangle_{\bf n}$ and of $\langle \lambda \rangle_{\bf n}$,
related to these initial conditions by the conditional averaging
procedure, are taken from the applicable solutions of Eq.\ (4).

It is difficult to guess {\it a~priori} when $\langle \lambda
\rangle_{{\bf n}}$ is negligible, because there is no small
parameter for the stretching part of $\lambda$. Therefore, let us
consider a generic set of conditions, presumably for the inertial
(convective) range, which can result in $\langle n_in_j
\partial v_i/\partial x_j \rangle_{{\bf n}} =0$. This can be a
combination of isotropy, which yields
$$
\langle n_in_j\rangle_{{\bf n}} =0 ~~~~(i\neq j) \eqno{(7)}
$$
and
$$
 \langle n_1^2\rangle_{{\bf n}}=\langle n_2^2\rangle_{{\bf n}}=
\langle n_3^2 \rangle_{{\bf n}}, \eqno{(8)}
$$
and directional randomness determined by equation
$$
\langle n_in_j \varphi \rangle_{{\bf n}} = \langle n_in_j
\rangle_{{\bf n}}\langle  \varphi \rangle_{{\bf n}}, \eqno{(9)}
$$
where $\varphi = \partial v_k/\partial x_l$ for arbitrary $k$ and
$l$. One should not mix the conditions (7)-(9) with global
isotropy and statistical independence of the velocity and
temperature gradients. In particular, (7)-(9) can be satisfied in
the inertial range even in the presence of strong global
anisotropy and correlation between gradients (see below for more
comments).

If we use conditions (7)-(9) in the presence of the
incompressibility condition $\partial v_i/\partial x_i=0$ we
obtain
$$
\langle \lambda \rangle_{{\bf n}} = -D \langle \left(
\frac{\partial n_i}{\partial x_j}\right)^2 \rangle_{{\bf n}}.
\eqno{(10)}
$$
That is, the formal difference between Eq.\ (2) for $T$ and the
conditionally averaged Eq.\ (6) for $G$ is reduced to the
"$\lambda$" term with the $\lambda$ given by Eq.\ (10). Equation
(6) can then be reduced, in Lagrangian variables, to
$$
\frac{dG}{dt}=\langle \lambda \rangle_{{\bf n}} G, \eqno{(11)}
$$
with the ``multiplicative noise" $\langle \lambda \rangle_{{\bf
n}}$ given by Eq.\ (10).

Weak diffusion of Lagrangian ``particles" can be described as
their wandering around the deterministic trajectories. The
introduction of a weak diffusion is equivalent to the introduction
of additional averaging in Eq.\ (11) over random trajectories
\cite{zeld1}. The small parameter $D$ in (10) and (11) will then
determine a slow time in comparison with the time scales in the
inertial range and will therefore not affect scaling properties of
$G$ in that range.

We should emphasize that the conditional average indicated by
$\langle \dots \rangle_{{\bf n}}$ is quite different from the
global average indicated by $\langle \dots \rangle$. Because of
this, the quantity $G$ in Eq.\ (6) remains a fluctuating variable.
To eliminate the stretching part from the conditionally averaged
coefficient $\langle \lambda \rangle_{{\bf n}}$, one does not need
to satisfy conditions (7)-(9) for all realizations of the
temperature gradient field ${\bf G}$, but only for the subset of
realizations that gives the main statistical contribution to the
spectrum of the magnitude $G$ in the inertial range. Therefore,
conditions (7)-(9) could well be violated globally without
affecting the main conclusion.

The essential point here is that the conditionally averaged
velocity $\langle {\bf v} \rangle_{{\bf n}}$ is smoothed
substantially in comparison with ${\bf v}$, while the fluctuation
of $G$ itself is still rapid in the diffusion-advection equation
(6) (because it remains in tact under the conditional average, by
virtue of its definition). Under these typical circumstances, the
natural expectation (see, for instance, Ref.\ \cite{ynaog} and
references therein) is that the spectral law for a quantity
governed by the diffusion-advection equation has a ``$-1$" region.
The result owes itself to the pioneering work of Batchelor
\cite{batc}, who applied this general idea to the
viscous-convection range of passive scalar fluctuations. While the
two contexts are quite different, they are the same in the sense
that the velocity field is smooth.

There is another way of deducing the $-1$ power law. We recall
that the spectral density fluctuations of temperature in this
region of scales has an approximately $-7/5$-ths slope. This slope
can be derived by the dimensional considerations used by Bolgiano
\cite{my}. We may apply the same reasoning to Eq.\ (6). Thus,
introducing an analogy of the dissipation rate for $G$, namely
$\chi_{\ast} = dG^2/dt$, Bolgiano's dimensional arguments yield
the scaling law for $G$ to be
$$
E_G (k) \sim \langle \chi_{\ast} \rangle^{4/5} (\alpha g)^{-2/5}
k^{-1} \eqno{(12)}
$$
in the inertial range. Figure 1 shows the corresponding spectrum
observed for a one-dimensional derivative of $G$ in our
experiment.
\begin{figure}\vspace{-0.4cm}
\centering \epsfig{width=.45\textwidth,file=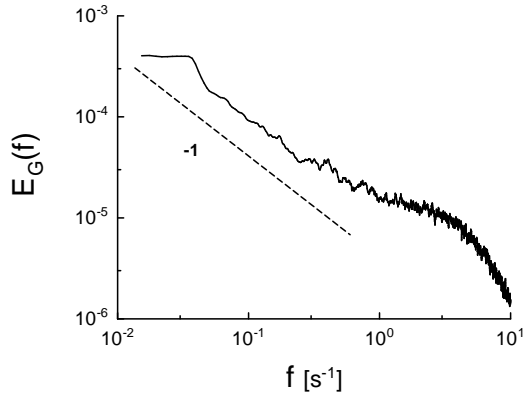}
\vspace{-5.1cm} \caption{Spectrum of a one-dimensional surrogate
of the magnitude of the temperature fluctuations gradient in
thermal convection. The straight line is drawn to indicate the
power-law spectrum (12).}
\end{figure}
For the effectively smoothed velocity field, the space
autocorrelation function can be characterized by a logarithmic
behavior \cite{chertkov} given by
$$
C(r)= \langle G(r) G(0) \rangle \sim \ln \left(\frac{L}{r}
\right), \eqno{(13)}
$$
or, using Taylor's hypothesis, in terms of $\tau$ by
$$
C(\tau )= \langle G(\tau) G(0) \rangle \sim \ln
\left(\frac{\tau_0}{\tau} \right). \eqno{(14)}
$$
This is seen from Fig.\ 2 to apply quite precisely for the data.
In our approach, the active character of the temperature in the
convection manifests itself through nontrivial properties of the
locally averaged velocity field in eq. (6). As a consequence, for
instance, the very significant input scale $L$ or $\tau_0$ for the
autocorrelation function has approximately the same size as
typical size of the largest plums observed in the convection. The
input scale can be readily calculated from Fig. 2 ($\tau_0 \simeq
10sec$).

\begin{figure}\vspace{-0.7cm}
\centering \epsfig{width=.45\textwidth,file=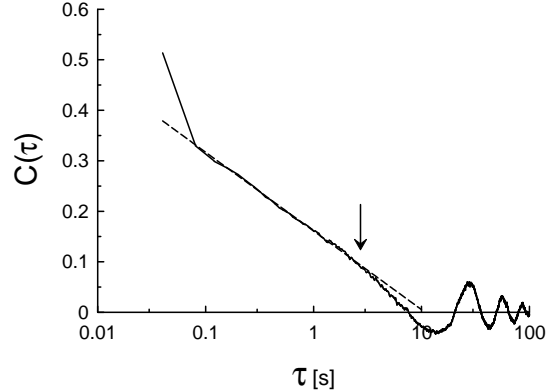}
\vspace{-4.8cm} \caption{Autocorrelation function $C(\tau)$
plotted against $\log \tau$, computed for the same data used for
the spectral calculations. The vertical arrow here and in Fig.\ 3
indicates the end of the inertial range.}
\end{figure}

The similarity of the $T$ and $G$ fields can also be seen for the
scaling of the dissipation rate itself. The local temperature
dissipation can be characterized by a gradient measure \cite{my}
as
$$
\chi (r) =\frac{\int_{\Omega_r} (\bigtriangledown{T})^2 dv}{v_r},
\eqno{(15)}
$$
where $\Omega_r$ is a subvolume with space-scale $r$. The scaling
law for the moments of this measure,
$$
\frac{\langle \chi (r)^p \rangle}{\langle \chi (r) \rangle^p} \sim
r^{-\mu_p},   \eqno{(16)}
$$
is an important characteristic of the dissipation intermittency
\cite{my, sa}. Using Taylor's hypothesis, one can define the local
dissipation rate as
$$
\chi (\tau) \sim \frac{\int_0^{\tau} (\frac{dT}{dt})^2 dt}{\tau},
\eqno{(17)}
$$
and the corresponding scaling of the moments of the local
dissipation rate \cite{sa} as
$$
\frac{\langle \chi (\tau)^p \rangle}{\langle \chi (\tau)
\rangle^p} \sim \tau^{-\mu_p}. \eqno{(18)}
$$

Analogous considerations can be brought to bear for the magnitude
of the temperature gradient
$$
\chi_{\ast} (\tau ) \sim \frac{\int_0^{\tau} (\frac{dG}{dt})^2
dt}{\tau}, \eqno{(19)}
$$
with exponents $\mu_p^*$. We show in Fig.\ 3 the dependence of the
normalized moments $\langle \chi (\tau)^p \rangle$ and $\langle
\chi_{\ast} (\tau)^p \rangle$ on $\tau$ calculated for the data
obtained in thermal convection.
\begin{figure}
\centering \epsfig{width=.45\textwidth,file=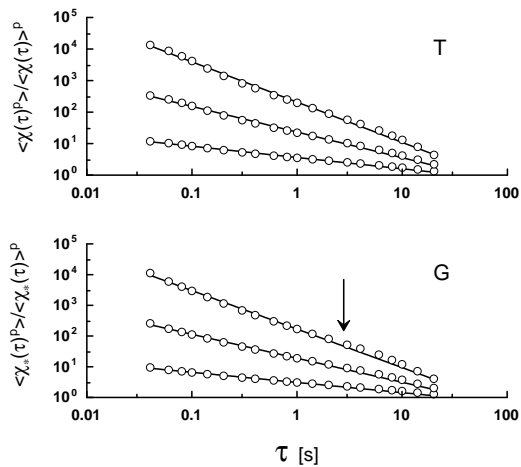}
\vspace{-4cm} \caption{Normalized moments $\langle \chi (\tau)^p
\rangle$ and $\langle \chi_{\ast} (\tau)^p \rangle$ against $\tau$
for the data obtained in convection ($p=2,3,4$). The straight
lines drawn to indicate scaling are best fits to the data on the
left of the vertical arrow.}
\end{figure}
The slopes of these straight lines provide us with the values of
the intermittency exponents $\mu_p$ and $\mu_p^*$, which are shown
in Fig.\ 4.
\begin{figure}
\centering \epsfig{width=.45\textwidth,file=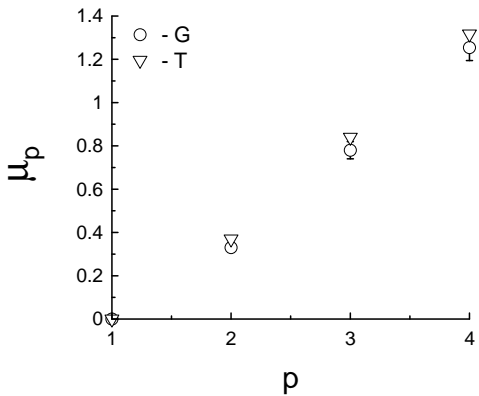}
\vspace{-5cm} \caption{Intermittency exponents $\mu_p$ and
$\mu_p^*$ extracted as slopes of the straight lines in Fig.\ 3
(equation (18)).}
\end{figure}
The two sets of intermittency exponents obtained for $T$ and $G$
are very close.

In summary, we have derived an equation for the magnitude of the
temperature gradient $G$ in thermal convection, and shown that
there are general circumstances under which the equation is
identical to that governing the temperature itself. The main
difference is that the velocity appearing the new equation, being
a conditional average, is a smoothed field. For the
advection-diffusion equation governed by a smooth velocity field,
it is natural to expect a power-law spectrum with a slope of $-1$;
measurements of the magnitude of the spectral density of $G$ are
consistent with this expectation. The correlation function of $G$
shows a logarithmic behavior, also as expected. Finally, the
scaling of the square of the derivative of $G$ has scaling
exponents that are identical to those of the temperature itself,
confirming that a deep analogy exists between $T$ and $G$ in the
inertial range.

We thank J.\ Schumacher and V.\ Yakhot for brief discussions at an
early stage of the work.

\end{document}